\documentclass[twocolumn,english]{revtex4}
\usepackage[T1]{fontenc}
\usepackage[latin9]{inputenc}
\usepackage{graphicx}
\usepackage{esint}

\makeatletter
\@ifundefined{textcolor}{}
{%
 \definecolor{BLACK}{gray}{0}
 \definecolor{WHITE}{gray}{1}
 \definecolor{RED}{rgb}{1,0,0}
 \definecolor{GREEN}{rgb}{0,1,0}
 \definecolor{BLUE}{rgb}{0,0,1}
 \definecolor{CYAN}{cmyk}{1,0,0,0}
 \definecolor{MAGENTA}{cmyk}{0,1,0,0}
 \definecolor{YELLOW}{cmyk}{0,0,1,0}
 }

\makeatother

\usepackage{babel}
\begin{document}

\title{Experimental evidence of the ferroelectric phase transition near the $\lambda-$point in liquid water. }

\author{P.O. Fedichev$^{1}$, L.I. Menshikov$^{2}$, G.S. Bordonskiy$^{3}$,
A.O. Orlov$^{3}$}

\affiliation{$^{1}$Quantum Pharmaceuticals, Kosmonavta Volkova 6A-606, Moscow,
125171, Russia}

\affiliation{$^{2}$Russian Research Center {}``Kurchatov Institute'', Kurchatov
Square 1, Moscow, 123182, Russia}

\affiliation{$^{3}$Institute of Natural Resources,Ecology and Cryology SB RAS,
Butina 26, Chita, 672000,Russia}
\begin{abstract}
We studied dielectric properties of nano-sized liquid water samples
confined in polymerized silicates MCM-41 characterized by the porous
sizes $\sim3-10nm$. We report the direct measurements of the dielectric constant
by the dielectric spectroscopy method at frequencies $25Hz-1MHz$
and demonstrate clear signatures of the second-order phase transition
of ferroelectric nature at temperatures next to the $\lambda-$ point
in the bulk supercooled water. The presented results support the previously
developed polar liquid phenomenology and hence establish its applicability
to model actual phenomena in liquid water.
\end{abstract}
\maketitle
Large dipole moments of individual molecules and the high density
of water are reasons behind the very rich phase diagram and quite
a few {}``anomalous'' properties \cite{debenedetti2003supercooled,su1998surface,iedema1998ferroelectricity,singer2005hydrogen,jahnert2008melting,morishige1999freezing,jackson1997thermally}.
Understanding the dielectric response of the liquid plays a crucial
role in modeling molecular interactions in computational physical
chemistry, biophysics, and drug design applications \cite{tomasi1994molecular,cramer1999implicit}.
The static dielectric constant $\epsilon$ of water at room temperatures
is very large, increases as the temperature decreases, and even diverges
in supercooled water being extrapolated to the unreachable temperature
$T_{C}\approx228^{0}K$ of the so called $\lambda$-transition \cite{hodge1978relative}.
Most other thermodynamic quantities, such as isothermal compressibility,
density, diffusion coefficient, and viscosity, are also singular \cite{angell1973anomalous,speedy1976isothermal,hodge1978relative,ter1981thermodynamic}.
Following the earlier idea of \cite{debye1929polar}, the authors
of \cite{angell1983sw,stillinger1977theoretical} indicated that the
phase transition could have ferroelectric features. The ferroelectric
hypothesis was also supported by a number of molecular dynamics (MD)
simulations \cite{wei1992flc,wei1993orientational,ponomareva2005atomistic}.
For example, a ferroelectric liquid phase was observed in a model
of the so called {}``soft spheres'' with static dipole moments \cite{wei1992flc,weis1992oos,groh1994fps,groh1994lro,weis1993ferroelectric,weis2005ferroelectric}.
The conclusion seems to be model independently confirmed in the MD
of hard spheres with point dipoles \cite{weis1992oos,matyushov2007model},
soft spheres with extended dipoles \cite{ballenegger2003sad}, and
in a model of fluids made of two-state particles with a nonzero dipole
in the excited state \cite{matyushov2005paraelectric}. 

Unfortunately, the relation of the ferroelectric phase transition
to the $\lambda-$point in actual water or even the very existence
of the paraelectric phase may be difficult to confirm in MD simulations
with finite number of atoms. There are quite a few reasons to it.
First of all due to the long range nature of the dipole-dipole interactions
between the molecules, the simulated liquid tend to form tightly correlated
domains and the calculated properties depend strongly on the boundary
conditions in any reasonably sized system \cite{weis2005ferroelectric}.
In the same time because of the low temperatures and the proximity
of various phase transitions the relaxation processes take a very
long time, which transforms into a necessity to simulate large and
strongly interacting molecular systems in realistic force-fields for
a very long time. 

The same problems naturally hinder theoretical understanding of the
physics behind the phase transition. There could be no purely electrostatic
MD model of the phase transition, since classical systems with electrostatic
interactions only are inherently unstable. Chemical forces, such as
hydrogen bonds, are known to play a very important role in water molecule
ordering at all temperatures in general, but particularly when close
to the phase transition \cite{bernal1933theory,pople1951molecular}.
The {}``minimal'' continuous model capable of predicting finer effects
depending both on the hydrogen bonding properties and the electrostatic
interactions of the water molecules was proposed in \cite{fedichev2008fep,fedichev2006long,men2011}.
One of its predictions is the ferroelectric phase transition in liquid
water within the temperature interval $T_{C}=-37\div-47^{0}C$, which
is remarkably close to the $\lambda-$point. In \cite{hodge1978relative}
the singularity of dielectric constant $\varepsilon$ as extrapolated
to the unreachable temperature $T_{C}=-45^{0}C$ was indeed reported.
On the other hand due to its weak character and unattainability of
$T_{C}$ the authors of \cite{hodge1978relative} did not attribute
it to the a ferroelectric phase transition. 

Fortunately the freezing temperature can be essentially lowered in
water confined in nanopores (see \cite{jahnert2008melting,morishige1999freezing,schreiber2001melting,petrov2009nmr}
and references therein). The porous systems can be studied by vapor
pressure measurements \cite{puri1957freezing}, dilatometry \cite{antoniou1964phase,litvan1966phase},
calorimetry \cite{antoniou1964phase,litvan1966phase,brun1977new,rennie1977melting,van1993very,takamuku1997thermal,hansen1997heat},
nuclear magnetic resonance \cite{rennie1977melting,takamuku1997thermal,hansen1997heat,puri1957freezing,pearson1974nmr,overloop1993freezing,stapf1995proton,ishizaki1996premelting,hansen1997pore,webber2004structural},
neutron diffraction \cite{takamuku1997thermal,steytler1985neutron,steytler1983neutron,dunn1988structural,bellissent1993structural},
small-angle neutron scattering \cite{li1991small,webber2004structural},
and $X$-ray diffraction \cite{jahnert2008melting}. The measured
transition temperature depression can be approximated as $\triangle T\approx-K/(R-t)$,
where $R$ is the characteristic radius of the pore, $K=52$K$\cdot nm$,
and $t=0.4nm$ is the (empirical) hydration layer thickness \cite{schreiber2001melting}.
Already for the pore radii below $R\sim2nm$ the decrease in freezing
temperature can be as large as $-50K$ and hence the predicted ordering
phase transition at $T=T_{C}$ can be observed and studied experimentally.
Moreover, in hindsight we believe that the first signatures of such
a transition were already observed in \cite{morishige1999freezing}.
In this work we present the results of the direct measurements of
the low-frequency liquid water dielectric constant, $\epsilon$, in
nanopores by dielectric spectroscopy method. The observed singularity
of $\epsilon$ near the $\lambda-$ point temperature in the supercooled
bulk water is a clear signatures of the second-order transition of
ferroelectric nature, both in quantitative and qualitative agreement
with the polar liquid phenomenology. 

The polar liquid phenomenology \cite{fedichev2006long} extends the
continuous models \cite{marcelja1976repulsion,Ramirez,gong2009langevin,azuara2008incorporating,koehl2009beyond,beglov1997integral,groh1994fps,frodl1992bai,kornyshev1983non}
and was originally developed to describe electrostatic energies of
biomolecules in aqueous solutions for drug discovery applications
\cite{joce2010application}. The model naturally describes the ordering
phase transition in water \cite{fedichev2008fep,men2011}. Within
the suggested model the polar liquid is characterized by the vector-field
$\mathbf{s}(\mathbf{r})=\left\langle \mathbf{d}\left(\mathbf{r}\right)\right\rangle /d_{0}$,
where $\mathbf{d}\left(\mathbf{r}\right)$ is the vector of the static
dipole moment of molecule residing at point $\mathbf{r}$, $d_{0}$
is its absolute value, $0<s(\mathbf{r})<1$. Total dipole moment of
molecule equals $\mathbf{d}_{t}\left(\mathbf{r}\right)=\mathbf{d}\left(\mathbf{r}\right)+\mathbf{d}_{e}\left(\mathbf{r}\right)$,
where $\mathbf{d}_{e}\left(\mathbf{r}\right)$ is the dipole moment
induced in the electronic shell. The Helmholtz free energy of polar
liquid is described by the functional of independent variables $\mathbf{s\left(\mathbf{r}\right)}$
and $\mathbf{d}_{e}\left(\mathbf{r}\right)$:
\[
F\left(\mathbf{s\left(\mathbf{r}\right)},\mathbf{d}_{e}\left(\mathbf{r}\right)\right)=P_{0}^{2}\int dV\left(\frac{C}{2}\sum_{\alpha,\beta}\frac{\partial s_{\alpha}}{\partial x_{\beta}}\frac{\partial s_{\alpha}}{\partial x_{\beta}}+V(\mathbf{s}^{2})\right)+
\]

\begin{equation}
+\int dVn_{0}^{2}\frac{2\pi\mathbf{d}_{e}^{2}}{\left(\epsilon_{\infty}-1\right)}+{\displaystyle \int dV\frac{1}{8\pi}\mathbf{E}_{P}^{2}}-\int dV\mathbf{P\left(\mathbf{r}\right)E_{\mathbf{e}}\left(\mathbf{r}\right)}\label{eq:FreeEnergyPhenom}
\end{equation}
Here $C$ is the phenomenological parameter responsible for the $H$-bond
network rigidity, $P_{0}=n_{0}d_{0}$, $n_{0}$ is the particle density
of the liquid, $\epsilon_{\infty}$ is the part of the dielectric
constant of the liquid which is unrelated to the molecular rotations
and accounts properly for the polarization of the internal (e.g. electronic)
degrees of freedom. It is the dielectric constant at frequencies larger
than that correspondent to reorientation of molecules. Next, $\mathbf{E}_{P}=-\mathbf{\nabla}\varphi_{P}$
is the polarization electric field produced by the polarization charges
with a density $\rho_{P}=-\mathbf{\nabla P}$ (sometimes it is called
as the depolarizing field), $\mathbf{E}_{e}\left(\mathbf{r}\right)=-\nabla\varphi_{e}$
is the external electric field induced by external charges with the
density $\rho_{e}\left(\mathbf{r}\right)$, $\mathbf{P}\left(\mathbf{r}\right)=P_{0}\mathbf{s}(\mathbf{r})+n_{0}\mathbf{d}_{e}\left(\mathbf{r}\right)$
is the polarization vector of liquid at point $\mathbf{r}$, $\mathbf{E}\left(\mathbf{r}\right)=\mathbf{E}_{P}\left(\mathbf{r}\right)+\mathbf{E}_{e}\left(\mathbf{r}\right)=-\nabla\varphi$
is the total electric field at point $\mathbf{r}$, $\varphi=\varphi_{P}+\varphi_{e}$.
Electric potentials $\varphi_{P}$ and $\varphi_{e}$ should be found
from correspondent Poisson equations $\Delta\varphi_{P}=-4\pi\rho_{P}$,
$\Delta\varphi_{e}=-4\pi\rho_{e}$. 

The phenomenological dimensionless function $V(s^{2})$ serves as
the polar liquid equation of state and can not be established in general
form. Its specific form can be found by comparing the model results
obtained with the help of Eq.(\ref{eq:FreeEnergyPhenom}) with the
results of MD simulations of a specific liquid. On the contrary, in
the small $s^{2}\ll1$ limit the function takes a nearly universal
form: $V(s^{2})\approx As^{2}/2+Bs^{4}$, $A=4\pi\tau/(3\epsilon_{\infty})$
has a universal (i.e. the same for any polar liquid) form, $B=3\pi/(5\epsilon_{\infty})$
is a liquid-specific constant, $B\sim1$, and $\tau=(T-T_{c})/T_{c}$,
where, at last, 
\begin{equation}
T_{c}=\frac{4\pi n_{0}d_{0}^{2}}{9\epsilon_{\infty}}\label{eq:Tc}
\end{equation}
is the critical temperature within the model. The physical picture
behind the change of the coefficient $A$ sign can be recovered from
the following argument. For uniformly polarized liquid, $\mathbf{s}(\mathbf{r})={\rm const}$,
the free energy (\ref{eq:FreeEnergyPhenom}) takes Landau-like form
\cite{LandauLifshitzStatPhys5}: 
\[
F=V\frac{P_{0}^{2}}{2}\left(\frac{2\pi\tau}{3\epsilon_{\infty}}s^{2}+Bs^{4}\right).
\]
At temperatures $T>T_{C}$ the equilibrium state correspond to the
disordered paraelectric phase with $\langle\mathbf{s}\rangle=0$,
whereas at lower temperatures $T<T_{C}$ the polar liquid undergoes
the second order phase transition and transforms to the long-range
ordered ferroelectric state. 

Second order ferroelectric phase transition should manifest itself
as a singularity in the liquid dielectric constant $\epsilon$. To
analyze the dielectric response we apply a uniform electric field
$\mathbf{E}$ and calculated the polarization of the liquid by the
minimization of free energy (\ref{eq:FreeEnergyPhenom}):
\begin{equation}
2V^{\prime}\left(\mathbf{s}^{2}\right)\mathbf{s}=\frac{1}{P_{0}}\mathbf{E},\:\mathbf{d}_{e}=\frac{\epsilon_{\infty}-1}{4\pi n_{0}}\mathbf{E}\label{eq: s in the uniform field}
\end{equation}
Consequently, in the weak electric field limit $E\ll P_{0}$ at $\left|\tau\right|\ll1$
the static dielectric constant of the liquid is given by: 
\begin{equation}
\epsilon=\epsilon_{\infty}\left(1+\frac{3}{\tau}\right),\: T>T_{C};\epsilon=\epsilon_{\infty}\left(1+\frac{3}{2\left|\tau\right|}\right),\: T<T_{C}.\label{eq:epsilonT}
\end{equation}
Therefore the measurements of the temperature dependence in $\epsilon(T)$
should exhibit a $\lambda$-point feature and diverge at $T=T_{C}$.
The exact value of the critical temperature for water can be obtained
using either of a few published measurements of the asymptotic values:
$\epsilon_{\infty}\approx4.9$ from \cite{stogrin1971}, $\epsilon_{\infty}=5.1$
from \cite{liebe1991model}, or $\epsilon_{\infty}=5.5$ from \cite{hasted1948dielectric}.
Respectively, Eq. (\ref{eq:Tc}) gives $T_{c}=236K$ ($-37{}^{0}C$),
$T_{c}=226K$ ($-47{}^{0}C$), and $T_{c}=210K$ ($-63{}^{0}C$).
All the numbers are remarkably close to $T_{c}\approx228K$ \cite{angell1973anomalous,speedy1976isothermal,hodge1978relative}
or $T_{c}\approx231K$ measured in supercooled bulk water \cite{ter1981thermodynamic}. 

\begin{figure}
\includegraphics[width=0.9\columnwidth]{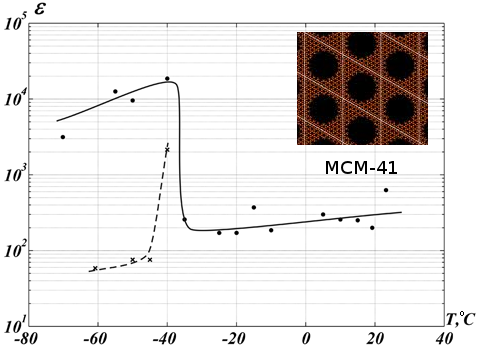}

\caption{Temperature dependence of low-frequency real part of dielectric permittivity
of liquid water (solid curve) and ice (dashed) in MCM-41 pores with
diameters $3.5$nm. \label{fig:Temperature-dependence-of water permittiviy}}
\end{figure}

To resolve the issue we investigated the dielectric response of the
water samples confined in polymerized silicate MCM-41 characterized
by a typical pore diameters of $D\sim3\div10$nm \cite{bordonsky2011,Kirik2009}.
We used the dielectric spectroscopy method at frequencies range 25Hz-1MHz.
The results of the measurements are summarized on Fig \ref{fig:Temperature-dependence-of water permittiviy}
and show a very distinguished $\lambda$-feature at $T_{C}^{exp}\approx-35^{0}C$
in full accordance the second-order phase transition picture presented
above. At first glance at large temperatures the data seem to suggest
a rise of $\epsilon$ with temperature in contradiction to both earlier
experiments \cite{hodge1978relative} and the theoretical prediction
(\ref{eq:epsilonT}). In fact the measurement errors in the dielectric
spectroscopy experiment are quite large and the contradiction disappears
entirely when the close vicinity to the transition point is considered.
Next to the observed transition temperature the singularity of $\epsilon$
is much stronger stronger than the previously reported dependence
$\epsilon\propto|\tau|^{-\alpha}$ characterized by the critical index
$\alpha\approx0.13$ \cite{hodge1978relative}. The mean field theoretical
prediction (\ref{eq:epsilonT}) $\alpha=1$ results in a much better
agreement with the measured values. Close to the phase transition
one can consider using a more more refined approach. Indeed, when
$\left|\tau\right|\ll1$, the fluctuations in the model (\ref{eq:FreeEnergyPhenom})
are {}``force''-less, $\nabla\cdot\mathbf{s}=0$, $\rho_{P}=0$,
$\mathbf{E}_{P}=0$ \cite{fedichev2006long}, the dipole-dipole interactions
vanish and the scale invariant calculation from \cite{patashinskii1979fluctuation}
gives $\alpha\approx1+\left(4-d\right)/6$, where $d$ is the number
spatial dimensions. Actually $d=3$ and $\alpha=7/6\approx1.2$, which
gives is stronger singularity than in the mean field theory. Though
it is hard to tell from a few points we have, the experimental divergence
of $\epsilon$ on Fig. \ref{fig:Temperature-dependence-of water permittiviy}
appears even stronger than each of the theoretical predictions. This
should not be surprising since our model is indeed oversimplified.
The calculations can be further improved by including at least the
two important forms of the liquid water states (such as the hexagonal
and the cubic water structures as explained in e.g. \cite{poole1992phase,stanley1994there}).
If we associate the ferroelectric phase transition with a single (say,
cubic \cite{stillinger1977theoretical}) component only, then the
critical behavior of the dielectric constant may be changed quite
dramatically by the sharp temperature dependence of the cubic water
fraction \cite{mallamace2008nmr}.

The hydration thickness $R_{D}\ll t\alt L_{T}$, where $R_{D}\approx0.2nm$,
and $L_{T}\approx0.7nm$ are characteristic phenomenological scales
of the liquid characterizing roughly the water molecules cluster and
the maximum correlated domain sizes respectively \cite{fedichev2008fep,fedichev2006long,men2011}.
Thus the pores in MCM-41 are very large compared to the hydration
layer thickness, $D\gg t,L_{T}$, and therefore the water in nanopores
still has all the properties of the bulk water. The equilibrium freezing
temperature of water in MCM-41 systems with $D=3.5$nm is $-38{}^{0}C<T_{C}^{exp}$
\cite{schreiber2001melting} and therefore the water samples in our
experiments are still liquid at temperatures $T\sim T_{C}^{exp}$
near the $\lambda$-point. The liquid and the solid water states were
definitely distinguished using the hysteresis effect \cite{jahnert2008melting,johari2009does,webber2004structural}:
the solid curve on Fig. \ref{fig:Temperature-dependence-of water permittiviy}
corresponds to the supercooled liquid state and the dashed line describes
the dielectric properties of the overheated ice state. 

In this way water freezing temperature suppression and the hysteresis
phenomena typical for water in nanopores let us actually investigate
the properties of the supercooled bulk liquid water in the temperature
range $-70\div+23{}^{0}C$. The dielectric spectroscopy measurements
reveal the strong singularity in dielectric constant at temperatures
corresponding to the position of $\lambda$-point of the bulk water.
Therefore the experimental data point support the earlier theoretical
predictions \cite{fedichev2008fep,men2011} of existence of the second
order phase transition of ferroelectric nature. Eq.(\ref{eq:Tc})
reconcile earlier theoretical predictions of \cite{weis2005ferroelectric}
with the experimentally observed values for the phase transition temperature
by introducing electronic polarizations $\epsilon_{\infty}$. The
very fact that the dielectric constant singularity was actually observed
suggests that experiments with water in nanoporous materials can be
used for in-depth studies of complicated phase diagram of water.

The authors wish to thank Quantum Pharmaceuticals and its team for
their support of the work and fruitful discussions. The experiments
reported in the work are part of Project No 22 of the SB Russian Academy
of Sciences. MCM-41 was provided by Dr. V.A. Parfenov and Dr. S.D.
Kirik.

\bibliographystyle{apsrev4-1}
\bibliography{../Qrefs}

\end{document}